\documentclass[prl,twocolumn,showpacs,amsmath,amssymb,superscriptaddress]
{revtex4}

\usepackage{graphicx}% Include figure files
\usepackage{dcolumn}% Align table columns on decimal point
\usepackage{bm}% bold math
\usepackage{hyperref}% add hypertext capabilities
\usepackage{color,ulem,subfigure}

\newcommand{\dr}[1]{|#1\rangle}
\newcommand{\dl}[1]{\langle#1}

\newcommand{\tb}[1]{\textbf{#1}}
\newcommand{\tr}[1]{\textrm{#1}}
\newcommand{\beq}{\begin{eqnarray}}
\newcommand{\eeq}{\end{eqnarray}}
\newcommand{\bem}{\begin{pmatrix}}
\newcommand{\eem}{\end{pmatrix}}
\newcommand{\f}{\frac}

\begin{document}

\preprint{APS/123-QED}

\title{Mass and chirality inversion of a Dirac cone pair in St\"{u}ckelberg interferometry}
\author{Lih-King Lim}
\affiliation{Laboratoire de Physique des Solides, CNRS UMR 8502, Univ. Paris-Sud, 91405 Orsay, France}
\affiliation{Laboratoire Charles Fabry, Institut d'Optique, CNRS, Univ. Paris-Sud, 2 avenue Augustin Fresnel, F-91127 Palaiseau cedex, France}
\author{Jean-No\"el Fuchs}
\affiliation{Laboratoire de Physique Th\'{e}orique de la Mati\`{e}re Condens\'{e}e, CNRS UMR 7600, Universit\'{e} Pierre et Marie Curie, 4 place Jussieu, F-75252 Paris, France}
\affiliation{Laboratoire de Physique des Solides, CNRS UMR 8502, Univ. Paris-Sud, 91405 Orsay, France}
\author{Gilles Montambaux}
\affiliation{Laboratoire de Physique des Solides, CNRS UMR 8502, Univ. Paris-Sud, 91405 Orsay, France}

\date{\today}

\begin{abstract}
We show that a St\"{u}ckelberg interferometer made of two massive Dirac cones can reveal information on band eigenstates such as the chirality and mass sign of the cones. For a given spectrum with two gapped cones, we propose several low-energy Hamiltonians differing by their eigenstates properties. The corresponding inter-band transition probability is affected by such differences in its interference fringes being shifted by a new phase
of geometrical origin. This phase can be a useful bulk probe for topological band structures realized with artificial crystals.
\end{abstract}

\maketitle

\textit{Introduction.} Topological properties of band structure are key to the modern classification of quantum phases of matter \cite{Volovik:03,Hasan:10}. In his seminal work, Haldane has shown that a pair of gapped Dirac cones realizing a trivial insulator can be turned into a Chern insulator upon reversal of the mass sign of a \textit{single} cone \cite{Haldane:88}. The resulting quantum anomalous Hall effect was recently measured in a magnetic topological insulator \cite{Chang:13}.
In addition to their mass sign, Dirac cones are also characterized as quantized vortices in the relative phase of their spinor eigenstates (corresponding to $\pm 1$ winding number or chirality) \cite{Fuchs:10, Park:11}.

Recent developments in artificial solids open the field of topological band structure engineering \cite{Shao:08,Goldman:10,Singha:11,Alba:11,Hauke:12,Cooper:12,Gomes:12,Bellec:13,Cooper:13,Polini:13}. Standard solid state techniques that are used to extract topological information - such as Shubnikov-de Haas oscillations, quantum Hall measurements and Landau-level spectroscopy - are typically unavailable in these systems. On the other hand, they offer the possibility of measuring new physical observables, such as that studied in cold atoms experiments \cite{Soltan:11,Aidelsburger:11,Tarruell:12,Price:12,Atala:12,Abanin:13,Wang:13,Goldman:13,Liu:13}. For instance, the long coherence time typical of cold atoms permits the study of St\"{u}ckelberg interferences in an optical lattice \cite{Kling:10,Lim:12}. In this Letter, we show that the phase in the St\"uckelberg interference pattern contains information not only on the energy bands \cite{Shevchenko:10}, but also on geometrical quantities characterizing the band eigenstates.
%------------------------------
\begin{figure}[ht]
\begin{center}
\includegraphics[width=8cm]{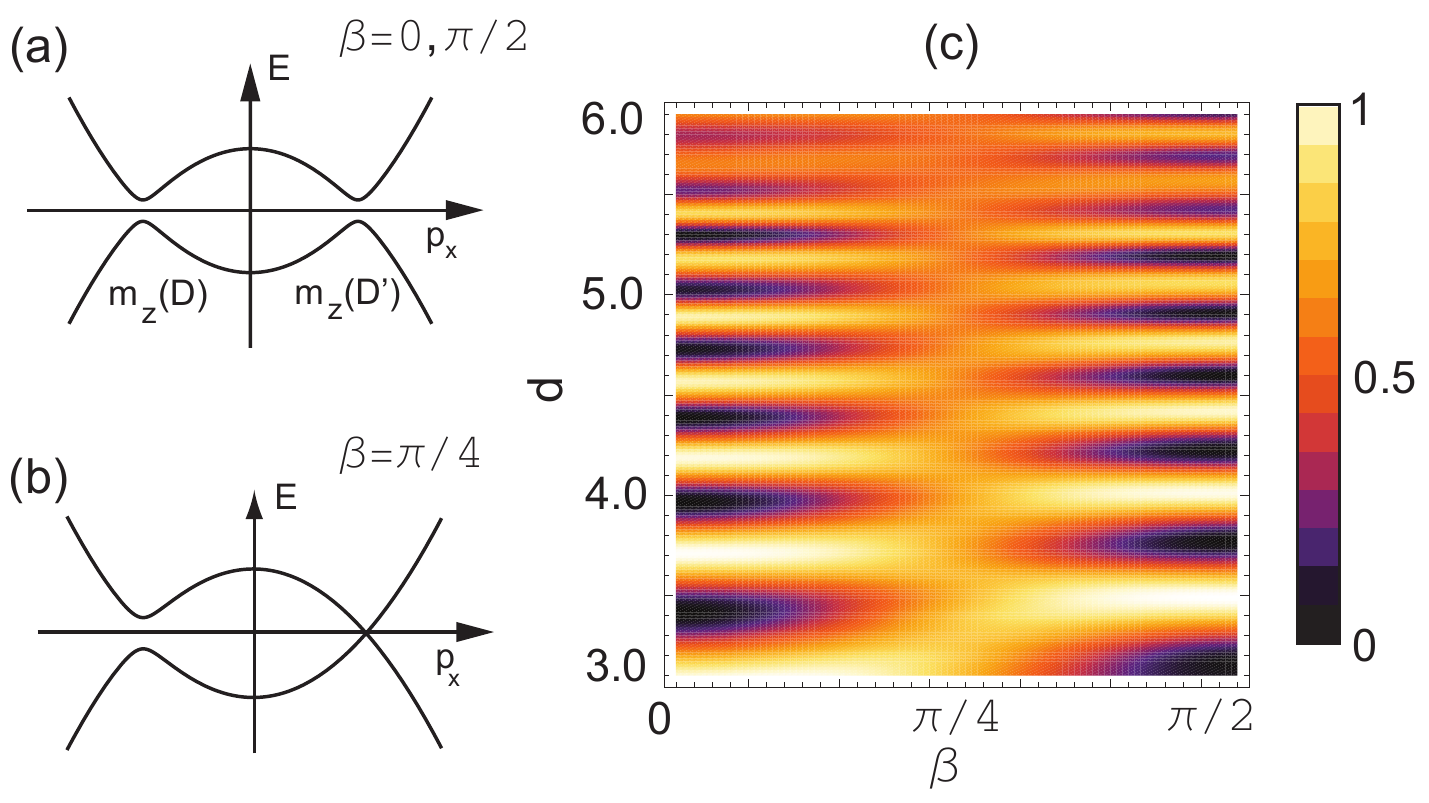}
\end{center}
\caption{St\"{u}ckelberg interferometer made of a pair of one-dimensional Dirac cones described by Hamiltonian (\ref{plusminus}) $H(p_x,p_y=0)$ with mass function $M_z(p_x)=M[\cos \beta-(p_x/\sqrt{2m\Delta_*})\sin \beta]$ tunable by the flux parameter $\beta \in [0,\pi/2]$. (a) $\beta=0$, $M_z(D)=M_z(D')$ (equal mass); $\beta=\pi/2$, $M_z(D)=-M_z(D')$ (opposite mass); (b) $\beta=\pi/4$, $M_z(D)\neq M_z(D')=0$ (single massless cone); (c) inter-band transition probability $P_f$, as a function of $\beta$ and the inter-cone distance $d=2\sqrt{2m \Delta_*}$, obtained with the replacement $p_x \to F_x t$ (see text) and $M=0.8(\hbar F_x)^{2/3}/(2m)^{1/3}$.}
\label{fig:smallk}
\end{figure}
%------------------------------

In order to illustrate our findings, we first consider a toy-model St\"uckelberg interferometer made of a pair of one-dimensional gapped Dirac cones a distance $d$ apart in reciprocal space (Fig.~\ref{fig:smallk}a). A particle initially in the lowest band moves from negative to positive momentum $p_x$ under the influence of a constant force and encounters the double cone structure. The two avoided crossings act as beam splitters controlled by Landau-Zener (LZ) tunneling \cite{Zener:32}. A ``flux'' parameter $\beta$ allows one to tune the relative sign of the two Dirac masses, similar to Haldane's model \cite{Haldane:88}. The latter can be realized in an optical lattice with cold atoms, see e.g. Ref. \cite{Shao:08,Alba:11}. At $\beta=0$, the two masses have the same sign and fringes are clearly seen in the final transition probability as a function of the distance between the cones (Fig.~\ref{fig:smallk}c). A mass inversion (induced by the parameter $\beta$ going from 0 to $\pi/2$), while keeping the bulk energy bands \textit{unchanged} (Fig.~\ref{fig:smallk}a), nevertheless leads to a $\pi$-shift in the St\"uckelberg interference fringes (Fig.~\ref{fig:smallk}c, compare $\beta=0$ and $\pi/2$). At the transition ($\beta=\pi/4$, Fig.~\ref{fig:smallk}b) one of the Dirac cones becomes gapless and the interference contrast fades.

The basic understanding of such a $\pi$-shift stems from the Berry phase of band eigenstates \cite{Berry:84,Shapere:89}. As the particle is accelerated through two crossings in succession, phase information related to band eigenstates is encoded into the probability amplitude during tunneling events. Geometrical characteristics of a gapped Dirac point, such as its chirality and its mass \cite{Sticlet:12}, are thus rendered observable in the interferometry thanks to non-adiabatic transitions.

In the following, we introduce several double LZ Hamiltonians corresponding to the same energy spectrum but differing by the chirality of Dirac cones, their relative mass sign and also consider different trajectories in reciprocal space. We first concentrate on a specific case, which we solve using analytical and numerical methods to show that the usual St\"uckelberg interferences in the inter-band transition probability  are shifted by what is shown to be a geometrical contribution. Then, we briefly consider all other cases for which we give an analytical expression of the geometrical phase shift.

\textit{Low-energy double cone Hamiltonian.} We define a class of effective two band models featuring two \textit{distinct} Dirac cones by the following two-dimensional Bloch Hamiltonian \cite{Montambaux:09}:
\beq\label{plusminus}
H(\vec{p})=\left(\f{p_x^2}{2m}-\Delta_*\right)\hat{\sigma}_x+ c_y p_y\hat{\sigma}_y+M_z(\vec{p})\hat{\sigma}_z.
\eeq
$\vec{p}=(p_x,p_y)$ is the momentum, $m$ gives the band curvature in the $x$ direction, $c_y>0$ is the $y$-direction velocity, $\Delta_*\geq 0$ is the merging gap \cite{changesign} -- which determines the distance $d$ between the two Dirac cones located at valleys $\vec{p}=D,D'\approx (\mp\sqrt{2m\Delta_*},0)$ -- $M_z(D)$ and $M_z(D')$ are the corresponding ``masses'' and $\hat{\sigma}_{x,y,z}$ are Pauli matrices operating in the pseudospin space. If the full Brillouin zone contains two and only two Dirac cones, the behaviour of the mass function $M_z (\vec{p})$ determines fully the quantum anomalous Hall state \cite{Hasan:10}. A constant mass function $M_z (\vec{p})=M$ describes equal Dirac masses and a vanishing Chern number. A mass function with a sign inversion in between $D$ and $D'$ for $M_z (\vec{p})=c_x p_x$, gives a non-zero Chern number in the individual bands.
The Hamiltonian (\ref{plusminus}) is therefore sufficiently general for describing different topological states. It describes Dirac cones with \textit{opposite} chirality in the two valleys. A second class of Dirac cones having the \textit{same} chirality can also be envisaged, as we discuss at the end of this Letter.

\begin{figure}[top]
\begin{center}
\includegraphics[width=8.5cm]{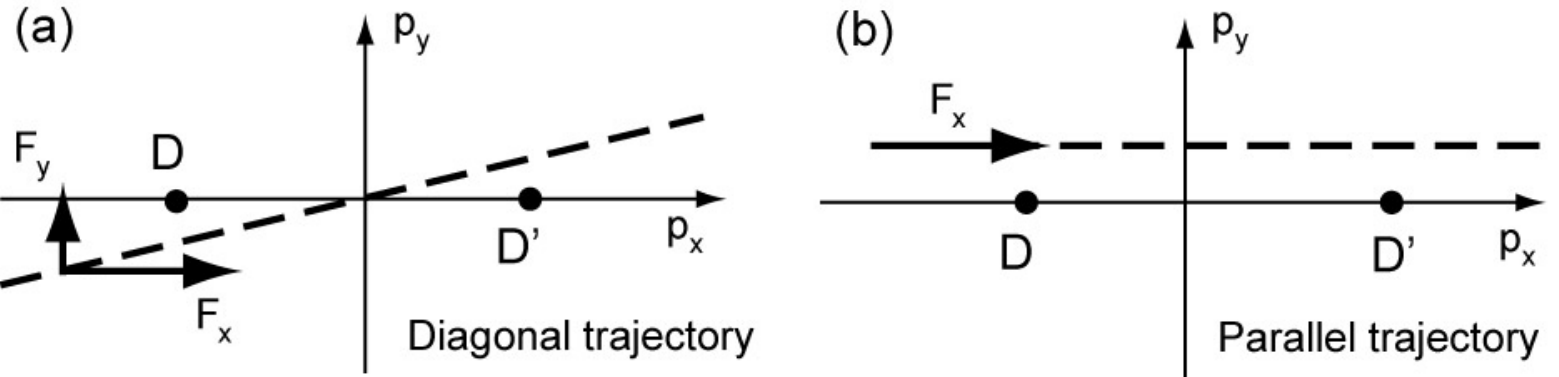}
\end{center}
 \caption{Momentum space trajectories driven by the force $\vec{F}=(F_x,F_y)$ in the vicinity of two Dirac points $D,D'$: (a) diagonal trajectory ($\vec{p}\to (F_x t,F_y t)$); (b) parallel trajectory ($\vec{p}\to (F_x t,p_y=\textrm{const.})$).}
 \label{fig:trajectories}
\end{figure}

\textit{Time-dependent Hamiltonian.} We now study inter-band dynamics of a particle experiencing a constant force $\vec{F}=(F_x, F_y)$ in such a system. The applied force is equivalent to a time-dependent gauge potential and thus leads to the substitution $(p_x,p_y)\rightarrow (p_{x}+F_x t,p_{y}+ F_y t)$ in the Bloch Hamiltonian (\ref{plusminus}). We distinguish two types of trajectories, one in which the two Dirac cones are on the opposite side of the full trajectory, termed ``diagonal" and meaning $\vec{p}\to (F_x t,F_y t)$, and the other in which the two Dirac cones are on the same side, termed ``parallel" and meaning $\vec{p}\to (F_x t,p_y=\textrm{const.})$, see Fig.~\ref{fig:trajectories}. We consider first the Hamiltonian with a constant mass function $M_z (\vec{p})=M$ and a diagonal trajectory, as the same consideration can be generalized to other cases (see Table \ref{summarytable}). The time-dependent Hamiltonian is
\beq\label{timeHam}
H(t)=c_y F_y t \hat{\sigma}_x +M \hat{\sigma}_y+ \left( \f{F_x^2 t^2}{2m}-\Delta_* \right)\hat{\sigma}_z
\eeq
where we have also performed a parameter-independent rotation in the pseudospin space $(\hat{\sigma}_x, \hat{\sigma}_y, \hat{\sigma}_z)\rightarrow (\hat{\sigma}_z, \hat{\sigma}_x, \hat{\sigma}_y)$. From now on, we use units such that $2m=F_x=\hbar=1$. The adiabatic spectrum is given by $E_{\pm}(t)=\pm\sqrt{c_y^2 F_y^2 t^2 +M^2+\left(t^2-\Delta_* \right)^2 }$ . By assuming  $\Delta_*\gg c_y^2 F_y^2 $, the two avoided crossings in the spectrum are located at $\pm t_0=\pm\sqrt{ \Delta_*}$. For an initial state in the lower band far from the first crossing, we seek to solve, in various limits, the probability for a particle ending up in the upper band after the second crossing. The most intuitive approach is to develop the so-called St\"{u}ckelberg theory, where the dynamics is assumed to be adiabatic except close to $t=\pm t_0$ where non-adiabatic transition occurs \cite{Shevchenko:10}. The adiabaticity parameter $\delta$ of the problem ``$\tr{gap}^2/ (\hbar\cdot \tr{force}\cdot \tr{speed})$" is given by $\delta=\Delta^2/(4 \sqrt{\Delta_*})$ with $\Delta=\sqrt{ \Delta_* c_y^2 F_y^2+M^2}$, and the St\"{u}ckelberg limit corresponds to the regime where the time separation $\sim 2 t_0$ between the two tunneling events is much larger than the tunneling time $t_{LZ}\sim \tr{max}(\sqrt{\delta},\delta)/\Delta$ \cite{Shevchenko:10}.

%------------------------------
%------------------------------
\textit{St\"{u}ckelberg theory.} We make a linear expansion in the Hamiltonian (\ref{timeHam}) around the crossings $t=-\xi t_0$ (where $\xi=\pm 1$ corresponds to the first/second crossing) to arrive at two LZ-type Hamiltonians
\beq\label{LZ}
H_{\xi}(t)=\bem -\xi\,2 t_0 t & -\Delta e^{i\varphi_\xi}\\
-\Delta e^{-i\varphi_\xi}& \xi\,2 t_0 t
\eem.
\eeq
The gap $\Delta e^{i\varphi_\xi}$ is generally complex with magnitude $\Delta=\sqrt{ \Delta_* c_y^2 F_y^2+M^2}$ and phase $\varphi_{+1}= \tan^{-1}[M/(c_y F_y \sqrt{\Delta_*})]$ and $\varphi_{-1}=\pi- \varphi_{+1}$, for $\xi=\pm 1$, respectively. It is crucial that the linearized Hamiltonian captures the full adiabatic spectrum $E_{\pm}(t)$ up to linear order in $t$ around the minima. In terms of the adiabaticity parameter $\delta$, the standard LZ formula $P_{LZ}=\exp(-2 \pi \delta)$ gives the tunneling probability of traversing one crossing \cite{Zener:32}. However, we are interested in the transition amplitudes where the phase information is also important. To this end, we express a general state in terms of the adiabatic basis of Hamiltonian (\ref{LZ}): $| \Psi (t)\rangle = b_+(t)|\psi_{+}(t)\rangle + b_-(t)|\psi_-(t)\rangle$ or in vectorial notation $\vec{b}(t)^T\equiv (b_+(t),b_-(t))$, and the basic element of the theory is first to construct the scattering $N_\xi$-matrix for each time crossing \cite{Shevchenko:10}. The $N_\xi$-matrix basically relates an asymptotic incoming-state $| \Psi (-t_a)\rangle$ to an asymptotic outgoing-state $| \Psi (t_a)\rangle$ across the crossing, i.e., $\vec{b}(t_a)=U(t_a,0^+)\cdot N_{\xi}\cdot U(0^-,-t_a) \cdot\vec{b}(-t_a)$ with the unitary evolution matrix $U(t_2,t_1)=\exp(-i\hat{\sigma}_z \int_{t_1}^{t_2}dt\, E_{+}(t))$ accounting for the dynamical phase, in the asymptotic time regime $t_a\gg \delta/\Delta$. Specifically, the time-dependent Schr\"{o}dinger equation for the Hamiltonian (\ref{LZ}) can be solved via Weber equation \cite{Zener:32}, giving
\beq
N_{\xi}=\bem \sqrt{1-P_{LZ}}e^{-i(\varphi_S+\xi\varphi_{\xi})}& -\xi\sqrt{P_{LZ}}\\ \xi\sqrt{P_{LZ}}& \sqrt{1-P_{LZ}}e^{i(\varphi_S+\xi\varphi_{\xi})}\eem.
\eeq
Except for the adiabatically accumulated dynamical phase, the $N_\xi$-matrices encode the rest of the information for the amplitudes across a single crossing. The Stokes phase $\varphi_S=\pi/4+\delta (\ln \delta-1)+\tr{arg}\Gamma(1-i\delta)$ is associated with the particle staying in the same band \cite{Shevchenko:10}. In addition, we find a non-perturbative correction $\varphi_\xi$-angle due to the phase of the complex gap.

To complete the St\"{u}ckelberg description for the full Hamiltonian (\ref{timeHam}) with two avoided crossings, we take the product of $N_\xi$-matrices -- one for each time crossing at $t=-t_0$ and $t_0$ -- and insert in between an unitary adiabatic evolution matrix $U(t_0,-t_0)$ to arrive at $N_{f}=N_{-1}\cdot U(t_0,-t_0)\cdot N_{+1}$. The transition probability $P_f$ in traversing two time crossings can be read off from the modulus squared of the matrix element $[N_{f}]_{12}$ giving
\beq\label{stuckprob}
P_{f}=4 P_{LZ}(1-P_{LZ})\, \sin^2(\varphi_{S}+(\varphi_{dyn}+\Delta \varphi)/2)
\eeq
with $\varphi_{dyn}\equiv \int_{-t_0}^{t_0}dt (E_+ -E_-)$ the dynamical phase and $\Delta \varphi\equiv \varphi_{+1}-\varphi_{-1}$. In the present case, $\Delta\varphi=-2\tan^{-1} [c_y F_y \sqrt{\Delta_*}/ M]$.
Eq.~(\ref{stuckprob}) has the usual St\"uckelberg structure $4 P_{LZ}(1-P_{LZ})\, \sin^2(\ldots)$ \cite{Shevchenko:10}, except that the interferences are shifted by an additional phase $\Delta \varphi$. The total phase consists of: (i) the Stokes phase $\varphi_S$, which only knows about the energy spectrum of a single avoided crossing through the adiabaticity parameter $\delta$; (ii) the dynamical phase $\varphi_{dyn}$, which contains information on the energy bands in between the two Dirac cones; and (iii) the phase shift $\Delta \varphi$, which is a new ingredient that results from the phase difference of the complex gap and encapsulates information on the band eigenstates. As we show below, this information is of geometrical nature -- as may already be guessed from its independence on $\delta$, i.e. on $\hbar$  -- and can be phrased in the language of a Berry phase, albeit involving two states, acquired along the path relating the two crossings.

\textit{Diabatic and adiabatic perturbation theory.} We now examine when the time-dependent problem defined by Hamiltonian (\ref{timeHam}) is amenable to perturbation theory in the diabatic $\delta\to 0$ and adiabatic $\delta\gg1$ limits, see Ref.~\cite{Fuchs:12} for similar notations. In the first case, the time evolution of the upper band amplitude $A_1(t)$ in the diabatic basis is given by
\beq
i  \dot{A}_1(t) \approx (c_y F_y t-i M) \exp \left[2 i\int^t dt' \left(t'^2-\Delta_*\right)  \right]
\eeq
with the lower band amplitude $A_2(t)\approx 1$. The equation can be integrated, with the boundary condition $A_1(-\infty)=0$, to give $A_1(+\infty)= -2^{2/3}\pi M  \tr{Ai} (s)- 2^{1/3} \pi c_y F_y \tr{Ai}\,'(s)$, with $s\equiv -2^{2/3}\Delta_*$ and $\tr{Ai}(s)$ is the Airy function. In the St\"{u}ckelberg limit, i.e., $t_0\gg t_{LZ}$, or equivalently, $|s|\gg 1$, we obtain $|A_1(+\infty)|^2\simeq 8 \pi \delta \sin^2 (\pi/4+(\varphi_{dyn}+\Delta \varphi)/2)$ which agrees with the diabatic limit ($\delta\to 0$) of expression (\ref{stuckprob}) with $P_{LZ}\simeq 1-2\pi \delta$, $\varphi_S\rightarrow \pi/4$, $\varphi_{dyn}\approx 8 \Delta_*^{3/2}/3 $ and $\Delta\varphi=-2\tan^{-1} [c_y F_y \sqrt{\Delta_*}/ M]$.

%------------------------------
\begin{figure}[top]
\begin{center}
\subfigure[]{\includegraphics[width=4cm]{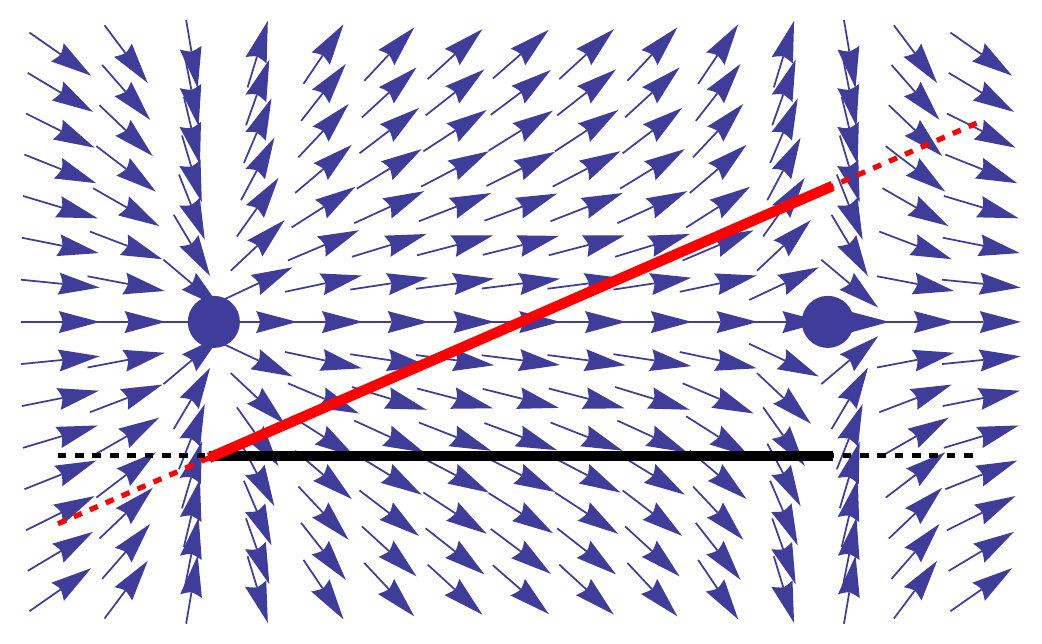}}
\subfigure[]{\includegraphics[width=4cm]{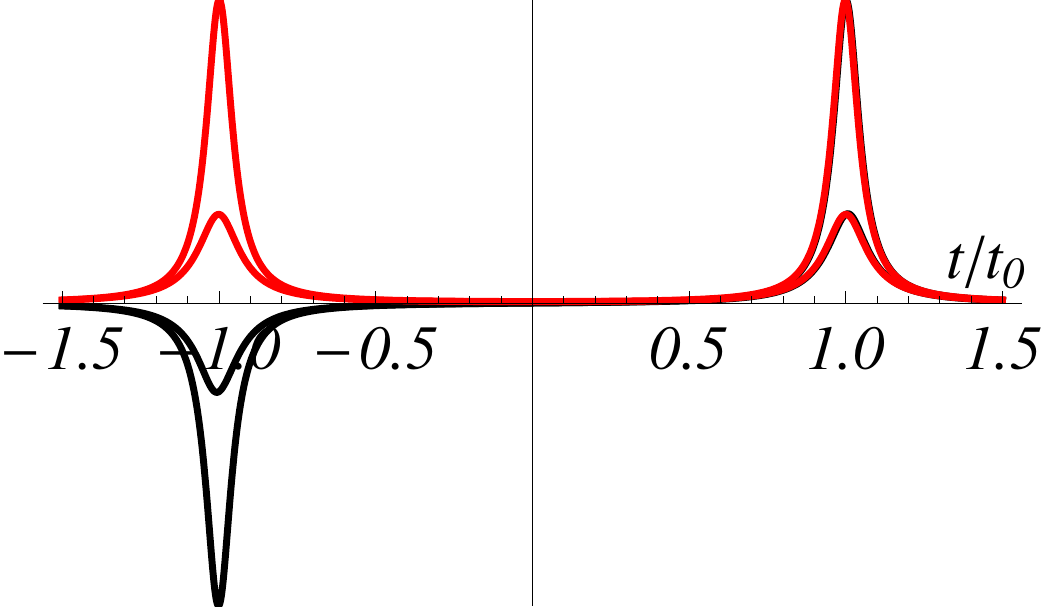}}
\end{center}
\caption{Hamiltonian (\ref{plusminus}) with $M_z(\vec p)=M$ is written as $H(\vec p)=\vec{B}(\vec p)\cdot \hat{\sigma}$ where the effective magnetic field $\vec{B}=E_+(\sin\theta\cos\phi,\sin\theta\sin\phi,\cos\theta)$ defines $\phi$ and $\theta$. (a) Vector plot of $\phi(\vec p)$ in the vicinity of a Dirac cone pair seen as a vortex and an anti-vortex indicated by dots. Parallel (black) and diagonal (red) trajectories are represented. (b) Berry connection $\dot \phi(t) ( 1 - \cos \theta(t))$ (in the south pole gauge \cite{gauge}) along the parallel (black) and diagonal trajectories (red) as a function of time $t/t_0$ in the St\"uckelberg limit $t_0\gg t_{LZ}$. Two values of the mass $M=0$ (higher peak) and $M \neq 0$ are considered. The geometrical phase $\varphi_g$ is obtained as a line integral of the Berry connection between $-t_0$ and $+t_0$  (thick lines in (a)).}
\label{fig:geomphase}
\end{figure}
%------------------------------
%--------------------------

In the adiabatic limit $\delta\gg1$ the time-dependent Schr\"{o}dinger equation in the adiabatic basis $\dr{\psi_{\pm}}$ gives
\beq
\dot{A}_{+}(t)\approx -\dl{\psi_+}|\partial_{t}\dr{\psi_-}e^{-i w(t)}
\eeq
where $A_-(t)\approx1$ and $w(t)\equiv \int^{t} d t' (E_- - E_+)- \int^{t} d t' \dl{\psi_-}|i \partial_{t'}\dr{\psi_-}+\int^{t} d t' \dl{\psi_+}|i \partial_{t'}\dr{\psi_+}$. Following Refs.~\cite{Dykhne:62}, the transition amplitude $A_+(\infty)$ can be obtained from the complex time crossings $E_+(t_{c})=0$, giving rise to two complex roots, $t_{c}$'s, lying closest to the real-time axis in the upper-half complex plane. The sum of the residue contributions leads to an interference effect and we find that the transition probability $|A_+(\infty)|^2 \propto \sin^2[(\varphi_{dyn}+\varphi_g)/2]$
where $\varphi_{dyn}$ is the dynamical phase introduced above, and $\varphi_g$ a gauge-invariant phase given by
 \begin{eqnarray}
\varphi_g &=& \int_{-t_0}^{t_0} d t\, \bigr( \dl{\psi_-}|i\partial_{t}\dr{\psi_-}-\dl{\psi_+}|i\partial_{t}\dr{\psi_+} \bigl) \nonumber \\
&+&\textrm{arg}\dl{\psi_-(-t_0)}\dr{\psi_-(t_0)} - \textrm{arg}\dl{\psi_+(-t_0)}\dr{\psi_+(t_0)}
\label{dthetageom}
\end{eqnarray}
which can be identified with a geometric phase for an open path involving two bands \cite{Gasparinetti:11} and a geodesic closure \cite{SB:88}. By comparing with  Eq. (\ref{stuckprob}) in the adiabatic limit ($\varphi_S\rightarrow 0$  when $\delta\gg 1$), the geometric nature of the phase shift $\Delta \varphi$ is revealed and we have precisely $\Delta \varphi=\varphi_g$.

In the specific case of Hamiltonian (\ref{timeHam}), we parameterize it as $H(t)=\vec{B}(t)\cdot \hat{\sigma}$ where $\vec{B}(t)=E_+(t)(\sin\theta(t)\cos\phi(t),\sin\theta(t)\sin\phi(t),\cos\theta(t))$ with $\tan\phi(t)=c_y F_y t/( t^2-\Delta_*)$, $\cos\theta(t)=M/E_+(t)$ in the unrotated $(\hat{\sigma}_x, \hat{\sigma}_y, \hat{\sigma}_z)$-basis. Then, the geometric phase is given by
$
\varphi_g=\int_{-t_0}^{t_0} dt \dot{\phi}(t)(1-\cos\theta(t))
$
in the south pole gauge \cite{gauge}. The integral can be evaluated to give $\varphi_g = -2\tan^{-1} [c_y F_y\sqrt{\Delta_*}/ M]$ in the St\"uckelberg limit, see Fig.~\ref{fig:geomphase}, identical to the phase shift $\Delta \varphi$.

\begin{table*}
\caption{Summary of phase shifts $\Delta \varphi$ realized for double Dirac cone interferometers differing by the chirality, relative mass sign and the trajectories. We assume well separated avoided crossings ($p_y^2\ll\Delta_*$ in the parallel, $F_y\ll 1$ in the diagonal and $c_x\ll 1$ in the opposite mass cases).}
\vspace{0.3cm}
\begin{tabular}{c|c|c|c}
\hline
Chirality & Mass function $M_z(\vec{p})$ (mass sign) & Parallel trajectory & Diagonal trajectory  \\
\hline
opposite & $M$ (identical)  & $0$ & $-2\tan^{-1} [c_y F_y \sqrt{\Delta_*}/M]$\\
opposite & $c_x p_x $ (opposite) & $2\tan^{-1} [c_x\sqrt{\Delta_*}/ (c_y p_y)]$ & $\pi$\\
identical & $M$ (identical) & $-2\tan^{-1} [2 p_y\sqrt{\Delta_*}/M]$ & $0$\\
identical & $c_x p_x $ (opposite) & $\pi$ & $2\tan^{-1} [c_x /(2F_y\sqrt{\Delta_*})]$\\
\hline
\end{tabular}
\label{summarytable}
\end{table*}
%-------------------------

%------------------------------
\begin{figure}[top]
\begin{center}
\includegraphics[width=6.5cm]{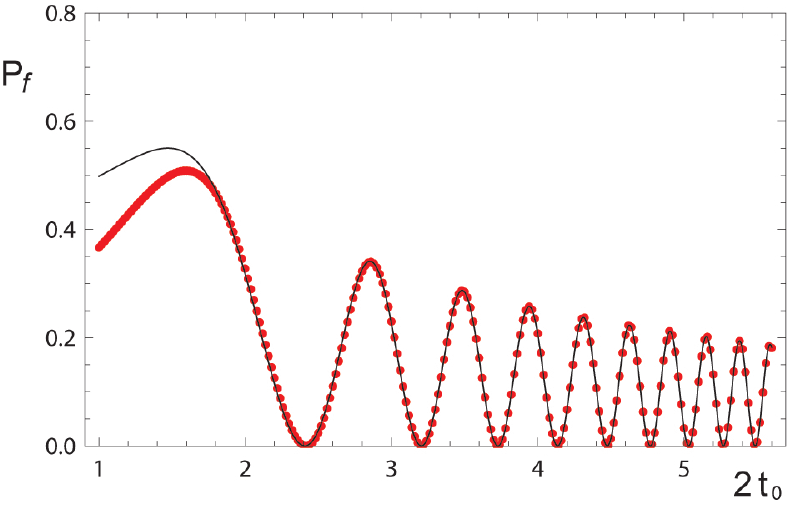}
\end{center}
\caption{Transition probability $P_f$ as a function of the time interval $2 t_0 = 2\sqrt{\Delta_*}$ between the two cones. Comparison between the St\"{u}ckelberg theory (Eq.~(\ref{stuckprob}), full curve) and numerical solution (dots) for the parameters: $c_y F_y=0.01$, $\Delta=0.3$.}
\label{fig:numerics}
\end{figure}
%------------------------------

\textit{Comparison with numerics.}
To verify our findings, we compare our results with numerical solutions of the time-dependent Schr\"{o}dinger equation expressed in both the diabatic and adiabatic bases, see Fig. \ref{fig:numerics}. We vary the path length by changing the separation $2 t_0$ between the two time crossings. We see that in the St\"{u}ckelberg limit, $t_0\gg t_{LZ}$, both the numerics and analytic results from Eq. (\ref{stuckprob}) are in good agreement.

\textit{Phase shift for general Hamiltonians.} Up to this point, we have mainly considered the case of a Dirac cone pair with \textit{opposite} chirality described by Hamiltonian (\ref{plusminus}), a constant mass and a diagonal trajectory. Using the same methods, we can compute the transition probability $P_f$ in three other cases corresponding to diagonal or parallel trajectories (as shown in Fig.~\ref{fig:trajectories}) and to equal ($M_z(\vec p)=M$) or opposite masses ($M_z(\vec p)=c_x p_x$). We always find that it satisfies Eq.~\ref{stuckprob}, albeit with a different phase shift $\Delta \varphi$ summarized in the first two lines of the Table~\ref{summarytable}.

It is also possible to consider a pair of gapped Dirac cones with the \textit{same} chirality given by
\beq
H(\vec{p})=\left(\f{p_x^2-p_y^2}{2m}-\Delta_*\right)\hat{\sigma}_x+ \f{p_xp_y}{m}  \hat{\sigma}_y+ M_z(\vec{p}) \hat{\sigma}_z\, ,
\label{plusplus}
\eeq
as occurs, e.g., around a single $K$ point in a twisted graphene bilayer \cite{Gail:11}. Changing the mass function and the trajectory type gives four additional cases, the phase shifts of which are given in the last two lines of Table~\ref{summarytable}.

\textit{Conclusion.}
The main result of our work is contained in Eqs.~(\ref{stuckprob}) and (\ref{dthetageom}) with $\Delta \varphi = \varphi_g$. These equations show that a St\"uckelberg interferometer carries information not only on the band energy spectrum but also on coupling between bands via a geometric phase. The latter could be accessed experimentally in a double cone interferometer involving Bloch oscillations and LZ tunnelings, e.g. with cold atoms in a graphene-like optical lattice as recently demonstrated with non-interacting fermions \cite{Tarruell:12}. The inter-band transition probability -- averaged over the initial Fermi sea -- can be measured in a time-of-flight experiment as the fraction of atoms that tunneled from the lower to the upper band during a single Bloch oscillation \cite{Tarruell:12,Lim:12}. Alternatively, a solid state realization of Bloch-Zener oscillations with multiple passages on a single Dirac cone has been proposed in a graphene ribbon superlattice, with constructive interferences showing up in the $I-V$ characteristics as sharp current peaks \cite{krueckl:12}. This can be generalized to the double cone case. In both realizations, a practical way of extracting the geometrical phase from the measured total phase of the interferometer is via the different force $F$ dependences: the dynamical phase scales as $1/F$, the Stokes phase varies slightly with $\delta\propto 1/F$, whereas the geometric phase is force independent \cite{foot2}.

\begin{acknowledgements}
We acknowledge useful discussions with F. Pi\'echon, U. Schneider and M. Schleier-Smith.
\end{acknowledgements}


\begin{thebibliography}{99}
\bibitem{changesign}Compared to \cite{Montambaux:09}, the merging gap is here defined with an opposite sign.
\bibitem{foot2} In the case of a diagonal trajectory $F_y/F_x$ is maintained constant as the total force $F=|\vec{F}|$ is varied.
\end{thebibliography}
\end{document}